# Abasy Atlas v2.2: The most comprehensive and up-to-date inventory of meta-curated, historical, bacterial regulatory networks, their completeness and system-level characterization

Juan M. Escorcia-Rodríguez[1], Andreas Tauch[2], and Julio A. Freyre-González[1,*]

[1]Regulatory Systems Biology Research Group, Laboratory of Systems and Synthetic Biology, Center for Genomic Sciences, Universidad Nacional Autónoma de México. Av. Universidad s/n, Col. Chamilpa, 62210. Cuernavaca, Morelos, México

[2]Centrum für Biotechnologie (CeBiTec). Universität Bielefeld, Universitätsstraße 27, 33615. Bielefeld, Germany

**\*Corresponding author:** jfreyre@ccg.unam.mx (JAFG)

## Abstract

Some organism-specific databases about regulation in bacteria have become larger, accelerated by high-throughput methodologies, while others are no longer updated or accessible. Each database homogenize its datasets, giving rise to heterogeneity across databases. Such heterogeneity mainly encompasses different names for a gene and different network representations, generating duplicated interactions that could bias network analyses. Abasy (**A**cross-**ba**cteria **sy**stems) Atlas consolidates information from different sources into meta-curated regulatory networks in bacteria. The high-quality networks in Abasy Atlas enable cross-organisms analyses, such as benchmarking studies where gold standards are required. Nevertheless, network incompleteness still casts doubts on the conclusions of network analyses, and available sampling methods cannot reflect the curation process. To tackle this problem, the updated version of Abasy Atlas presented in this work provides historical snapshots of regulatory networks. Thus, network analyses can be performed at different completeness levels, making possible to identify potential bias and to predict future results. We leverage the recently found constraint in the complexity of regulatory networks to develop a novel model to quantify the total number of regulatory interactions as a function of the genome size. This completeness estimation is a valuable insight that may aid in the daunting task of network curation, prediction, and validation. The new version of Abasy Atlas provides 76 networks (204,282 regulatory interactions) covering 42 bacteria (64% Gram-positive and 36% Gram-negative) distributed in 9 species (*Mycobacterium tuberculosis, Bacillus subtilis, Escherichia coli, Corynebacterium glutamicum, Staphylococcus aureus, Pseudomonas aeruginosa, Streptococcus pyogenes, Streptococcus pneumoniae*, and *Streptomyces coelicolor*), containing 8459 regulons and 4335 modules.

**Database URL:** https://abasy.ccg.unam.mx/

**Background**

Regulation at the gene transcription level is a fundamental process for bacteria to adapt to different media conditions and to cope with adverse environments. Transcription factors (TFs) mainly mediate this process. They are proteins capable to promote or hinder the transcription of their target genes (TGs). A TF-coding gene and its TGs conform a regulon, multiple regulons can be assembled to construct a gene regulatory network (GRN) where nodes and edges depict genes and interactions, respectively. Given the different specificity across TFs, they can contribute to organism adaptation in different levels which provides hierarchical and modular properties to GRNs in bacteria [1].

The increasing number of experimental strategies to study the transcriptional machinery [2] has allowed the community to unveil novel regulatory interactions. Despite curation efforts, many interactions remain buried in publications and are not integrated into a GRN yet. Organism-specific databases offer expertise and often are the primary resource for further research on the organism of interest. Such databases include RegulonDB [3] for *Escherichia coli*, DBTBS [4] and SubtiWiki [5] for *Bacillus subtilis*, CoryRegNet [6] for *Corynebacterium glutamicum* and MtbRegList [7] for *Mycobacterium tuberculosis*. Nonetheless, many of those databases are no longer updated or accessible [8]. Besides, the availability of multiple organism-specific databases gives rise to heterogeneity, which could bias results when cross-organisms analyses are performed. Such heterogeneity encompasses different names for the same gene and different network representations. This is even a problem for a single organism when complementary databases are integrated.

The analysis of global properties through multiple bacteria have revealed similarities among them [9-14]. Nonetheless, those studies have been limited to only a few organisms and results need to be validated with the most complete GRNs [15]. Besides, the study of the effect of network incompleteness on network structural analyses has been hindered by the limitations in databases to identify when a set of novel interactions is reported, and the experimental evidence supporting those interactions. Since no GRN curation model has been developed, works to study this phenomenon have been limited to simulate the curation process by decomposition or reconstruction of the GRNs by different random models [16,17].

Diverse databases cope with information inconsistency, such as CollecTF [18] for experimentally-validated TF binding sites in bacteria, and GSDB [19] for 3D chromosome and genome topological structures. Other resources integrating and homogenizing experimentally-validated data with computational predictions include STRING [20] for protein-protein interaction networks, SwissRegulon [21] for regulatory sites in prokaryotes and eukaryotes organisms, PRODORIC [22] for DNA binding sites for prokaryotic TFs, RegNetwork [23] for transcriptional and posttranscriptional regulatory relationships for human and mouse, and Network Portal (http://networks.systemsbiology.net/) for coregulation networks. But, poor efforts have been carried out to provide consolidated, disambiguated, homogenized high-quality GRNs on a global scale, their structural properties, system-level components, and their historical snapshots to trace their curation process.

Abasy Atlas v1.0 was originally conceived to fill this gap by making a cartography of the functional architectures of a wide range of bacteria [12]. Our database provides a comprehensive atlas of annotated functional systems (hereinafter also referred to as modules), statistical and structural network properties, and system-level elements for reconstructed and meta-curated (homogeneous and disambiguated) GRNs across 42 bacteria, including pathogenically and biotechnologically relevant organisms. Abasy Atlas is the first database in providing predictions of global regulators, basal machinery genes, members of functional modules, and intermodular genes based on the system-level elements predicted for the natural decomposition approach (NDA) in several bacteria [9,11-13]. The NDA is a biologically motivated mathematical approach leveraging the global structural properties of a GRN to derive its architecture and classify its genes into one of the four above-mentioned categories of system-level elements. Abasy Atlas was also designed to provide statistical and structural properties characterizing the GRNs, such as their associated power laws, percentage of regulators, network density and giant component size, and the number of feedforward and feedback motifs among others.

In this work, we present the expanded version of Abasy (**A**cross-**ba**cteria **sy**stems) Atlas, which consolidates information from different sources into historical snapshots of meta-curated GRNs in bacteria. Each historical snapshot represents the integrated knowledge we had about a GRN at a given time point. The new Abasy Atlas v2.2 makes possible to study the effect of network incompleteness across bacteria on diverse GRNs analyses, to identify potential bias and improvements, and to predict future results with more complete GRNs. Besides, Abasy Atlas GRNs integrates regulation mediated by regulatory proteins, small RNAs, sigma factors and regulatory complexes to better understand the biological systems [24]. This global representation of the GRNs eases their use because the organism-specific databases usually represent each network in a different file and different format, which can convolute the parsing of the network flat files and the integration of information.

While most proteins regulate gene transcription as homodimeric complexes, the regulation of gene expression can also be achieved by heteromeric complexes, whose subunits are encoded by different genes. Despite previous integrative approaches merging different level components [25-27], heterodimeric complexes have not been properly represented in most of them nor databases. One of the most common representations is to assign the regulations to each subunit, leading to a duplicated representation of the interaction in the GRNs. The new Abasy Atlas v2.2 provides a homogeneous representation for heteromeric complexes, when information is available, preserving the regulatory information and avoiding duplicated, misleading interactions.

In summary, Abasy Atlas v2.2 provides historical snapshots of reconstructed and meta-curated GRNs across bacteria, their completeness level, topological properties, and system-level components, enabling network completeness-dependent analyses for multiple organisms. Besides, the homogeneity of gene symbols, interactions confidence level, and network representation allow Abasy Atlas GRNs to be used as gold standards for benchmarking purposes, such as those to assess GRN predictions and theoretical models. In the section "Functionality", we describe studies that would be benefited from the functionality of Abasy Atlas v2.2 [28-35].

Abasy Atlas does not intend to replace organism-specific databases containing regulatory interactions with biological information such as regulatory sites. Conversely, it fills a gap by providing a consolidated version of bacterial GRNs on a global scale, their structural properties, system-level components, and their historical snapshots to trace their curation process. Abasy Atlas is cross-linked to diverse external databases providing biological, genomic, and molecular details. Cross-links to organism-specific databases included as a source for each GRN are also provided. From there, the user can further inquire about biological considerations such as binding sites annotation, TF conformation, genome annotation, and chromosomal conformation. All essential data when studying the molecular mechanisms and evolution of GRNs in bacteria. In this way, Abasy Atlas serves as an across-organisms database coping with information inconsistency and providing high-quality GRNs on a global scale.

Remarkable uses of previous versions of Abasy Atlas [12] comprise the characterization of *C. glutamicum* GRN [13], the integration of gene regulatory interactions to metabolism to identify the relevant TGs suitable for strain improvement [36], and comparative genomic analyses to characterize the transcriptome profile of *Corynebacterium pseudotuberculosis* in response to iron limitation [37]. Abasy Atlas v2.0 was used to identify evolutionary constraints on the complexity of GRNs enabling the study of three models to predict the total number of genetic interactions [14]. The latter allowed to compute an interaction coverage as a proxy of network completeness, which improves the biased network genomic coverage (fraction of the genome in the network). Abasy Atlas V2.2 could be useful to improve these works since more complete GRNs provide more information regarding transcriptional regulation in medically and biotechnologically relevant organisms such as *M. tuberculosis* and *C. glutamicum*. Also, to improve models developed with the previous version of Abasy, such as the novel network completeness model presented in the section "Estimating GRNs completeness by leveraging their constrained complexity".

**A primer on the natural decomposition approach: predicting global regulators, modular genes shaping functional systems, basal machinery genes, and intermodular genes**

Abasy Atlas was designed to provide annotations of the modules and system-level elements integrating each GRN. These predictions are computed by using the NDA. The NDA is a large-scale modeling approach characterizing the circuit wiring and its global architecture. It defines a mathematical-biological framework providing criteria to identify the four classes of system-level elements shaping GRNs: global regulators, modular genes shaping functional systems, basal machinery genes, and intermodular genes. Studies have shown that regulatory networks are highly plastic [38]. Despite this plasticity, by applying the NDA our group has found that there are organizational principles conserved by convergent evolution in the GRNs of phylogenetically distant bacteria [11]. The high predictive power of the NDA has been proven in previous studies by applying it to the phylogenetically distant *E. coli* [9], *B. subtilis* [11], and *C. glutamicum* [13], and by comparing it with other methods to identify modules [39].

The NDA defines objective criteria (e.g., the κ-value to identify global regulators) to expose functional systems and system-level elements in a GRN, and rules to reveal its functional architecture by controlled decomposition (Supplementary Figure 1). It is based on two biological premises [10,11]: (1) a module is a set of genes cooperating to carry out a particular physiological function, thus conferring different phenotypic traits to the cell. (2) Given the pleiotropic effect of global regulators, they must not belong to modules but rather coordinate them in response to general-interest environmental cues.

According to the NDA, every gene in a GRN is predicted to belong to one out of four possible classes of system-level elements, which interrelate in a non-pyramidal, three-tier, hierarchy shaping the functional architecture [10-13] as follows (Supplementary Figure 2): (1) Global regulators are responsible for coordinating both the (2) basal cell machinery, composed of strictly globally regulated genes and (3) locally autonomous modules (shaped by modular genes), whereas (4) intermodular genes integrate, at the promoter level, physiologically disparate module responses eliciting combinatorial processing of environmental cues.

## Construction and content

### Abasy Atlas current content

Abasy Atlas v2.2 provides the most complete set of experimentally curated GRNs across bacteria. Abasy Atlas represents regulatory interactions by using network models where nodes represent genes or regulatory protein complexes, and directed links depict regulatory interactions. Since the release of Abasy Atlas v1.0 in 2016 [12], the number of GRNs has increased from 50 to 76 (+52%) covering 42 bacteria (64% Gram-positive and 36% Gram-negative) distributed in 9 species (*Mycobacterium tuberculosis*, *Bacillus subtilis*, *Escherichia coli*, *Corynebacterium glutamicum*, *Staphylococcus aureus*, *Pseudomonas aeruginosa*, *Streptococcus pyogenes*, *Streptococcus pneumoniae*, and *Streptomyces coelicolor*) and 41 strains (Figure 1A and Supplementary Figure 3).

These 76 GRNs comprise 204,282 regulatory interactions (+160%) organized into 8,459 (+128%) regulons and 4,335 modules (+144%). We homogenized the representation of heteromeric TFs and their subunits and obtained a total of 12 heteromeric TFs, all of them in the GRN of *E. coli* K-12. However, this paves the way for a homogeneous representation of GRNs that will be propagated to more organisms in a future version of Abasy Atlas, when information regarding heteromeric TFs for these organisms is available. A total of 20 historical snapshots for the model organisms *M. tuberculosis*, *B. subtilis*, *E. coli,* and *C. glutamicum* were also included in the Abasy Atlas v2.2.

### Unique machine-readable, user-friendly identifiers for each GRN reconstruction

Studies using GRNs from organism-specific databases usually cite the source database. However, while some articles specify the GRNs used [28,39], others do not [9,40]. This drives to a reproducibility problem when the database updates the GRN and does not provide the historical snapshots. To cope with this problem, a machine-readable and user-friendly identifier was assigned to each network to ease reporting and identification when using the database.

Network identifiers are constructed as follows: Five fields are separated by an underscore, three are mandatory and two are optional. The first field represents the NCBI taxonomy ID of the organism (mandatory). The second field, preceded by a "v", which stands for version, is the year when the network was reconstructed (mandatory). The field starting with an "s" provides information about the sources from which the network was reconstructed (mandatory). The confidence level of the evidence supporting the regulatory interactions is described by an optional field starting with an "e". When this field is omitted means that the reconstruction contains all the available interactions disregarding the confidence level of evidence, whereas "strong" is used for those GRNs reconstructed only with interactions validated by direct experimental evidence. An optional description field, preceded by a "d", enables to include keywords such as "sRNA" for GRNs containing sRNAs-controlled regulons (Figure 1B).

The source field, that starting with an "s", is composed by a database name abbreviation and year when meta-curated from databases, and the last two digits of the publication year when curated from literature (see Supplementary Table 1 for a complete list of data sources abbreviations and references). On the "Browse" page of Abasy Atlas, the user can identify the source for each GRN, as well as for the subnetworks when the GRN is a meta-curation from different sources.

**Historical snapshots of the GRNs**

Network theory-based approaches to study the organizing principles governing GRNs have been pointed to be biased by the curation process and incompleteness [16,41]. Nevertheless, those studies have been mainly applied to subnetworks sampled by different random computational algorithms that cannot reproduce faithfully the curation process by the scientific community. To bring an alternative solution to this problem, we have been curating organism-specific databases and literature during the construction of Abasy Atlas in different time points for several organisms (hereinafter referred to as historical snapshots). Namely, nine historical snapshots for *E. coli,* four for *C. glutamicum,* four for *B. subtilis,* and three for *M. tuberculosis* (Figure 2).

Each historical snapshot represented in Figure 2 is the most complete version of the GRNs at that time point. However, individual GRNs are also available. For example, the historical snapshot of the GRN of *B. subtilis* in 2017 (224308_v2017_sDBTBS08-15-SW18, Figure 2) integrates regulatory interactions from two organism-specific databases (DBTBS [42] and SubtiWiki [5]) and one article [43] (Figure 3). The individual GRNs are available with their own network ID (224308_v2008_sDBTBS08_eStrong, 224308_v2017_sSW18, and 224308_v2015_s15, respectively). Note that the GRN from DBTBS is also the first historical snapshot for *B. subtilis* (Figure 2), and GRNs from different sources do not need to be from the same year since a new historical snapshot integrates every previous GRNs. The network integration and homogenization from different sources enables cross-bacteria analyses with the historical snapshots.

We will continue querying organism-specific databases and curating literature periodically to obtain more complete versions of each GRN. Also, we will extend the historical snapshots to other organisms as information will be available.

**Meta-curation of GRNs: Quality control coping with inconsistency and preserving information from the different sources**

The heterogeneity in gene symbols and network representations often conduces to redundancy and loss of information. Consequently, this heterogeneity can result in misleading network reconstructions. The meta-curation process mainly consists of homogenizing gene symbols and network representation before merging interactions from different sources. To cope with gene symbols disagreement among regulatory datasets from different sources, we gathered gene name, locus tag, and synonyms for each gene in the GRNs. Then, we developed an algorithm to map gene symbols onto unambiguous canonical gene names and locus tags. This allowed us to remove a total of 223 redundant nodes and 412 redundant interactions from the current set of GRNs (Supplementary Figure 4). We refer the reader to version 1.0 of Abasy Atlas for further information about the gene symbols disambiguation algorithm [12]. For the graphical network representation, we use the unambiguous canonical gene name when available or locus tag. This eases to identify genes of interest. However, the mapping of gene identifiers allows the user to use the search box with different gene symbols and synonyms mapping to the same gene and navigate through the neighborhood of the gene of interest.

Abasy Atlas also provides the confidence level supporting each interaction since GRNs composed with different confidence-levels may bias their structural properties [14]. Therefore, a "strong" or "weak" confidence level is assigned to each interaction according to an expanded scheme based on the one proposed by RegulonDB [44,45]. The basic idea of the confidence level scheme is to label as "strong" only those interactions with direct, non-ambiguous experimental support such as DNA binding of purified TF [45]. Besides, the meta-curated networks that merge regulons from different sources also integrate the effect and the evidence level. This makes the GRNs from Abasy Atlas the most complete collection of homogenous versions in contrast to those individual GRNs available in organism-specific databases.

One of the main caveats of consolidating networks is the non-machine readable, heterogeneous way to represent the information about the way a TF regulates a specific TG and the evidence supporting such interaction, mainly for community-updated databases. To tackle this problem, we manually curate those attributes from different sources when available. Thus, Abasy Atlas makes possible to know in a homogenous fashion whether a TF promotes or hinders its TGs transcription even for interactions from a community-updated database such as SubtiWiki. Therefore, if the same interaction from a different source share effect but diverge on evidence, the interaction and the "strong" evidence is conserved since one directly experimentally validated interaction is enough to classify the edge as "strong" [45]. On the other hand, in case of different effects and the same evidence level, both effects are conserved in a single dual interaction to avoid redundancy. In the case that both attributes are different, only the "strong" interaction is conserved (Supplementary Figure 5). This meta-curation process allows us to reconstruct the most complete GRNs available preserving information from the different complementary sources (Figure 3).

**Meta-curation of GRNs: Quality control filtering spurious interactions by reassessing the confidence level of each interaction**

We perform a meta-curation process to reduce the number of spurious interactions, thereby reassessing the confidence level of the interactions. Although networks with "weak" evidence are a valuable resource to study the transcriptional regulation, only directly experimentally validated interactions offer the reliability needed to use GRNs as gold standards. Abasy Atlas eases the selection of gold standards for benchmarking purposes through ready-to-download filtered "strong" GRNs (Supplementary Figure 6).

Using the historical snapshots of the *E. coli* GRNs, we analyzed how often a regulatory interaction identified by a "weak" methodology was validated as "strong" evidence. We found that the number of interactions identified for each methodology varies in a wide range, as well as its fraction of predictions validated as "strong" (Figure 4A). Namely, "inferred computationally without human oversight" (ICWHO) is the evidence with the lowest fraction of validated interactions (Figure 4A and Supplementary Figure 7). On the other hand, "RNA-polymerase footprinting" (RPF) is the only methodology having a 100% of interactions validated as "strong" evidence, and >50% of "gene expression analysis" (GEA) predictions have been validated despite being the "weak" evidence with the highest number of predictions.

We further analyzed the effect of the interactions with ICWHO as its unique evidence, and found that most of these interactions were present in the 2013 and 2014 time points but no longer in 2015 or later. Being this the reason for the outstanding completeness of these network reconstructions and its unusual system-level elements proportions (Figure 4B). For this reason, we decided to exclude predictions being supported only by the ICWHO evidence in Abasy Atlas. This analysis highlights the capability of the system-level properties to assess GRNs quality. It is important to note that despite the small fraction of validated interactions inferred by "non-traceable author statement" (NTAS) (Supplementary Figure 7), we did not remove interactions supported only by this evidence since the number of predicted interactions is very small (Figure 4A).

**Estimating GRNs completeness by leveraging their constrained complexity**

The ability to quantify the total number of interactions in the complete GRN of an organism is a valuable insight that will leverage the daunting task of curation, prediction, and validation by enabling the inclusion of prior information about the network structure. Besides, the ability to track the completeness, quantified as the fraction of the known interactions from the total number in the complete network (interaction coverage), through different historical snapshots could allow to develop models on how new regulatory interactions are discovered and to provide a framework to assess network analysis and network inference tools. But, poor efforts have been directed towards the longstanding problem of how to assess the completeness of these networks. Traditionally, network genomic coverage has been used as a proxy of completeness. The genomic coverage of a regulatory network is the fraction of genes in the network relative to the genome size. Nevertheless, this measure poses potential biases as it neglects regulatory redundancy and the combinatorial nature of gene regulation, thus potentially overestimating network completeness.

For example, the addition of a global regulon or sigmulon (perhaps discovered by high-throughput methodologies) to a quite incomplete regulatory network could bias the genomic coverage. Assume you have a regulatory network with a genomic coverage of 15% (600 / 4000) and 700 interactions. You then found a paper reporting the promoter mapping for the corresponding housekeeping sigma factor, whose sigmulon has 3000 genes (400 of which were already in the original network). Next, you found that 100 out of the 3000 interactions in the global sigmulon already exist in our original network. You then integrate all the remaining 2900 new interactions to your original network to found that your resulting network has a new genomic coverage of 80% (3200 / 4000) and 3600 interactions. This new high genomic coverage may suggest a highly complete network but it is indeed the same quite incomplete original network plus a single global sigmulon. To clarify this, assume that the total number of interactions in the complete network is 10000, then the completeness of this new network is 36% (3600 / 10000). Whereas the curation of a single housekeeping sigmulon increased the completeness ~30% (3600 / 10000 - 700 / 10000), the new completeness is still low, and the genomic coverage is highly overestimating when is used as a proxy for completeness. Therefore, to state the completeness of a regulatory network correctly, it is fundamental to estimate the total number of interactions. Two recent works have simultaneously provided estimations on the size of GRNs [14,46].

On one hand, the RegulonDB team carried out an exploratory analysis [46]. They used a single version of the *E. coli* regulatory network and high-throughput datasets of binding experiments for around 15 TFs. By assuming a linear model, they found an upper-bound estimate of 45759 regulatory interactions. They claimed that only one-third of the ~46000 would affect gene expression, concluding that the complete network comprises only around 13000 interactions.

Alternatively, our group recently explored the constraints on several structural properties of the 71 regulatory networks deposited in Abasy Atlas v2.0 [14]. We found that the network density ($d$) as a function of the number of genes ($n$) follows a power law as $d \sim n^{-\gamma}$ with $\gamma \approx 1$. Since 1972, a seminal paper by Robert May showed that the frontier between dynamical stability and instability for a complex system follows a power law as $d \sim n^{-1}$, relating complexity quantified via the density of interactions and the number of variables (the size of the system) [47]. The density of interactions (network density) is the fraction of potential interactions that are real interactions, thus a constraint in network density implies a constraint in the total number of interactions in the complete network. As we found that density is constrained in GRNs, we explored three possible models to predict the total number of interactions as a function of the number of genes (see Figure 4 in [14]): edge regression (assuming linearity, $R^2$ = 0.90), density invariance (assuming an invariant density, $R^2$ = 0.86) and density proportionality (assuming an exponential decay, $R^2$ = 0.91). All the models had a good fit to the data (0.86 ≤ $R^2$ ≤ 0.91), with small differences between them. These models predicted that the total number of interactions in the complete *E. coli* regulatory network is ~10000, ~14000, and ~11000, respectively.

After publication, we reformulated the problem. As regulatory networks are directed and self-regulations are allowed, the maximum number of possible interactions ($I_{max}$) is $n^2$ as each of the $n$ genes could regulate to other $n$ genes including itself (self-regulation). The density of a regulatory network must be then computed as

$$d = \frac{I}{I_{max}} = \frac{I}{n^2}$$

By introducing this equation into the power law found for the density of the Abasy Atlas networks ($d \sim n^{-\gamma}$), we derived another power law modeling the total number of interactions in the regulatory network as a function of the number of genes as

$$I = dn^2 \sim n^{-\gamma}n^2 \sim n^{2-\gamma}$$

This model has a better fit to data (Figure 5, $R^2$ = 0.98) than the previous three models, and allows us to compute the total number of interactions in the regulatory network of an organism as $I_{total} \sim$ (genome size)$^{2-\gamma}$. We implemented this model in Abasy Atlas v2.2 to provide estimations on the completeness of each regulatory network, including confidence intervals. The power-law model predicts that the complete *E. coli* regulatory network will have 11,656 total regulatory interactions. This model can learn the tendency in the number of interactions, and it improves as more regulatory networks are included in Abasy Atlas. That is one of the reasons motivating us to continue expanding Abasy Atlas by adding new organisms and historical snapshots.

**Homogeneous representation for heteromeric transcription factor complexes**

Even though heterodimeric regulatory complexes are not overrepresented in regulatory networks, some of them are global regulators and their interactions control up to ~10% of the genome and represent a valuable percent of the whole network (~6% in *E. coli* GRNs). IHF is a global regulator histone-like protein of *E. coli* that regulates transcription as a heterodimeric complex that is shaped by two different proteins: IhfA and IhfB. Although both subunits can form homodimeric complexes, the affinity for DNA is much lower [48], and no regulation in such fashion has been reported. For this reason, assigning the regulatory activity to each subunit (a gene-gene representation, Figure 6B) is a misleading representation. Additionally, the RpoS sigma factor allows the transcription of both subunits conforming IHF, which in turn also regulates its subunits (Figure 6A). Such interesting autoregulation cannot be properly represented in a gene-gene based representation (Figure 6B). Conversely, a representation of the IHF heteromeric complex regulating *ihfA* and *ihfB* is better as it depicts the IHF conformation and links them to the TFs regulating their transcription.

This representation is also useful for subunits of heteromeric regulatory complexes that can exhibit regulation in a homodimeric fashion, such as the *relB* product regulating *relE*, *hokD*, and its transcription both as a homodimer and as part of the RelBE complex with *relE* (Figure 6C). This RelE-RelB toxin-antitoxin system in *E. coli* [49] is not properly represented in a gene-gene network (Figure 6d) as it shows regulatory activity by the *relE* product on its own. This representation eases the application of the networks as gold standards for inference methods such as those based on the DNA sequence and TF binding sites prediction. For analysis requiring GRNs composed only by genes, Abasy Atlas provides the required information to identify the classification of each biological entity (Supplementary Figure 8). Currently, Abasy Atlas comprises 12 heteromeric TFs, all of them in the meta-curated GRN of *E coli* K-12 obtained from RegulonDB [46]. Future development includes the addition of heteromeric TFs in those organisms where this information is available.

**Updates for model organisms**

*Corynebacterium glutamicum* **ATCC 13032**

The PubMed database was screened to find papers published between January 2017 and August 2018 and describing new transcriptional regulatory interactions of *C. glutamicum*, in addition to the comprehensive data set previously deposited in Abasy Atlas [13]. Four new regulators of different types have been examined in detail, exerting in total 63 new direct transcriptional interactions. Moreover, the predicted regulatory role of the AraC/XylR-type protein Cg2965 (PheR) has been confirmed by experimental data [50,51]. PheR activates the expression of the *phe* gene (*cg2966*) encoding phenol hydroxylase, allowing *C. glutamicum* to degrade phenol by a meta-cleavage pathway. Electrophoretic mobility shift assays (EMSAs) demonstrated a direct interaction of the purified PheR protein with the *phe* promoter region [51]. The MarR-type regulator CrtR (Cg0725) is encoded upstream and in divergent orientation of the carotenoid biosynthesis operon *crtEcg0722crtBIYEb* in *C. glutamicum*. DNA microarray experiments revealed that CrtR acts as a repressor of the *crt* operon. Additional EMSAs with purified CrtR showed that CrtR binds to a region overlapping the −10 and −35 promoter sequences of the *crt* operon [52].

The two-component system EsrSR (Cg0707/Cg0709) controls a regulon involved in the cell envelope stress response of *C. glutamicum* [53]. Interestingly, the integral membrane protein EsrI (Cg0706) acts as an inhibitor of EsrSR under non-stress conditions. The resulting three-component system EsrISR directly regulates a broad set of genes, including the *esrI-esrSR* locus itself, and genes encoding heat shock proteins (*clpB*, *dnaK*, *grpE*, *dnaJ*), ABC transporters and putative membrane-associated or secreted proteins of unknown function. Among the target genes of EsrSR is moreover *rosR* (*cg1324*) encoding a hydrogen peroxide-sensitive transcriptional regulator of the MarR family and playing a role in the oxidative stress response of *C. glutamicum* [53,54].

The extracytoplasmic function sigma factor SigD (Cg0696) is a key regulator of mycolate biosynthesis genes in *C. glutamicum* [55]. Chromatin immunoprecipitation coupled with DNA microarray (ChIP-chip) analysis detected SigD-binding regions in the genome sequence, thus establishing a consensus promoter sequence for this sigma factor. The conserved DNA sequence motif 5′-GTAAC-$N_{17(16)}$-CGAT-3′ was found in all ChIP-chip peak regions and presumably corresponds to the −35 and −10 promoter regions recognized by SigD. The *rsdA* (*cg0697*) gene, located immediately downstream of *sigD*, is under direct control of a SigD-dependent promoter and encodes the corresponding SigD anti-sigma factor [55].

The WhcD protein (Cg0850) interacts with WhiA (Cg1792) to exert jointly an important regulatory effect on cell division genes of *C. glutamicum* [56]. WhiA is an exceptional transcriptional regulator as it has been classified as a distant homolog of homing endonucleases that retained only DNA binding activity [57]. Binding of the WhcD-WhiA complex to the promoter region of the cell division gene *ftsZ* was observed by EMSAs using purified fusion proteins, although WhcD alone did not bind to the genomic DNA. The sequence motif 5′-GACAC-3′ was found to be important for binding of the WhcD-WhiA complex to the DNA. Additionally, loss of the DNA-binding activity of WhiA in the presence of an oxidant indicated a regulatory role for this protein to control cell division of *C. glutamicum* under oxidative stress conditions [56].

We merge these interactions with the previous version of the GRN for *C. glutamicum* and included as a new historical snapshot (196627_v2018_s17) with 2317 genes (73.8% of genomic coverage) and 3444 interactions (45.8% of interaction coverage) (Figure 2). The "strong" version of the network was also included, containing a total of 2237 genes (71.3% of genomic coverage) and 2969 interactions (39.5% of interaction coverage).

*Mycobacterium tuberculosis* H37Rv

Chauhan et al. [58] reported 41 experimentally validated interactions among sigma factors and transcribed genes in the human pathogen *M. tuberculosis*. These interactions were added to the most recent *M. tuberculosis* GRNs and deposited in Abasy Atlas. The regulations among the sigma factors and TGs constitute a valuable contribution to the understanding of how *M. tuberculosis* sigma factors regulate their expression and therefore, their cellular concentrations to compete for the available RNA polymerases. Historical snapshots for the years 2015, 2016, and 2018 are available so far (Figure 2).

*Bacillus subtilis subtilis* **168**

Interactions from the most recent big update of SubtiWiki [5] were merged with the last version of Abasy Atlas including interactions from DBTBS [4] and a non-database hosted publication [43]. The result represents a new time point in the *B. subtilis* GRN history. Until now, four historical snapshots are available for this representative Gram-positive organism (Figure 2), being the last one the GRN with the highest genomic coverage in Abasy Atlas.

*Escherichia coli* **K-12 MG1655**

RegulonDB [46] is one of the first organism-specific databases for transcriptional regulation data and it continues being updated. This makes *E. coli* the organism with a higher number of historical snapshots. Meta-curated GRNs from 2003 to 2018 depict the effect of the curation process in this Gram-negative model organism (Figure 2). The meta-curation of the GRNs in Abasy Atlas reassesses the confidence level of the interactions (see "Construction and content"), and integrates the regulations by TFs, sRNAs, and sigma factors from RegulonDB into a global regulatory network.

**Utility and discussion**

**User interface**

From the "Home" page you can find the description and statistics of Abasy Atlas, as well as links of interest. In the "Browse" page you can find the species for which a global GRN is deposited in Abasy Atlas, along with the number of items (networks) for such species. Further, you can click on the species to identify the strains available and even the confidence level you need. After the selection of the strain and the confidence level, you will find the historical snapshots available for the GRN of interest, as well as additional information such as the genomic and interaction coverage, data sources, and fraction of the system-level components predicted by the NDA (Supplementary Figure 9). By clicking on "Global properties" you will find statistical and structural properties characterizing the GRN of interest. Such properties include the number of transcription factors, network density, size of the giant component, number of feedforward and feedback motifs, among others. On the same page, you can find the plots for degree, out-degree and clustering coefficient distributions (Supplementary Figure 10). We fitted these distributions to a power-law using robust linear regression of log-log-transformed data with Huber's T for M-estimation. This overcomes the negative effect of outliers, in contrast to ordinary least squares, which is highly sensitive to outliers in data.

You can directly search for a specific gene in the upper-right box from any page. Once you are visualizing the subnetwork of interest, using the interactive panel (Supplementary Figure 11) you can customize the visualization with several buttons and download the subnetwork as a high-definition PNG image, as well as the JSON file. Every global network can be downloaded from the "Downloads" page (Supplementary Figure 6). Regulatory networks are provided in JSON data-interchange format, including NDA predictions and, when available, effect and evidence supporting regulatory interactions. JSON is an open standard file format, which is a lightweight, language-independent, widely used, data-interchange format supported by >50 programming languages (e.g., Python, R, Matlab, Perl, Julia, JavaScript, PHP) through a variety of readily available libraries. JSON uses human-readable text to store and transmit data objects consisting of attribute-value pairs and array data types. The JSON data files downloadable from Abasy Atlas are readily importable into Cytoscape for further analyses. Gene information and module annotation flat files in tab-separated-value file formats are also available for download. Information on how to parse the JSON files is available in the "Downloads" page. The citation policy, and the methodology to identify the system-level elements and to predict the interaction coverage is available in the "About" page. You can find additional help on the "Help" page, and contact us on the "Contact" page for any subject; we will appreciate your feedback.

**Functionality**

Following, we describe some remarkable cases where this new version of Abasy Atlas could have been applied to improve the studies:

The DREAM5 consortium assessed to identify the best methodology to predict GRNs from gene expression data [28] using *E. coli* and *Staphylococcus aureus* as prokaryotic models. However, they did not study how its assessment was affected by network incompleteness. This analysis can be carried out by using the set of the historical snapshots for model organisms as gold standards. The same could be applied for other assessments such as identifying the best tools to predict TF binding sites [29], DNA motifs [29,30,59], and functional modules [31].

Further, Abasy Atlas could be used to extend those benchmarking studies to include more organisms. For example, DREAM5 considered only *E. coli* as a prokaryotic model to compute the overall score because a sufficiently large set of experimentally validated interactions for *S. aureus* did not exist at that time [28]. Currently, Abasy Atlas provides GRNs for 13 *S. aureus* strains, being USA300/TCH1516 the most complete one with 25 and 30.6% of genomic and interaction coverage, respectively.

In addition to benchmarking improvements, the comprehensive atlas of GRNs that Abasy Atlas provides could be applied to study the communication that exists between the regulation of gene transcription with other mechanisms such as protein-protein interactions and metabolism [32-34]. Even when only the regulation of gene transcription is studied, across-organisms information provided by Abasy Atlas can be used to trace the evolution of the GRN in bacteria, and compare them using gene orthology and network alignment [35]. Future development of Abasy Atlas includes GRNs comparative analyses based on their structural properties.

**Future development**

Despite high-throughput strategies to study transcriptional regulation, there is a lack of novel interactions reported in contrast with earlier years (Figure 2). Besides, only a handful of organisms have been experimentally studied. Computational approaches have been a hopeful option for non-model organisms and a plethora of algorithms to infer GRNs have emerged. Nonetheless, many of them are based solely on statistical approaches lacking biological constraints to filter spurious interactions. Previous assessments of tools to infer GRNs have unveiled their poor performance but also have shed light on the possibility to increase precision by consensus approaches and biological constraints [28].

Future development of Abasy Atlas aims to include inferred non-model organisms GRNs in a conservative fashion by different consensus-based approaches and the application of currently available data to validate predicted networks by using GRN organizing constraints, such as the composition of system-level elements (Figure 4B) and network structural properties. The addition of heteromeric TFs for more organisms is also considered in the short-term future development. Mainly for the model organisms *C. glutamicum* and *B. subtilis* for which more information regarding regulation by heteromeric TFs is available. Besides, historical snapshots for non-model organisms already available in Abasy Atlas, such *Streptomyces coelicolor* will be included, while continuing including additional historical snapshots for model organisms curated from the literature and organism-specific databases. Finally, a python library providing an API to allow programmatic access to Abasy Atlas, and a REST API are under development.

**Conclusions**

Beyond the regulon level, Abasy Atlas provides the most complete and reliable set of GRNs for many bacterial organisms, which can be used as the gold standard for benchmarking purposes and training data for modeling and network prediction. Besides, Abasy Atlas provides historical snapshots of regulatory networks. Therefore, network analyses can be performed with GRNs having different completeness levels, making it possible to identify how a methodology is affected by the incompleteness, to pinpoint potential bias and improvements, and to predict future results. Additionally, Abasy Atlas is the first database providing estimations on the completeness of GRNs, their global regulators, modules, and other system-level components. The estimation of the total number of regulatory interactions a GRN could have is a valuable insight that may aid in the daunting task of network curation, prediction, and validation. Furthermore, the prediction of the system-level elements in GRNs has allowed unraveling the complexity of these networks and provides new insights into the organizing principles governing them, such as the diamond-shaped, three-tier, hierarchy unveiled by the NDA. The GRNs in Abasy Atlas have been meta-curated to avoid heterogeneity such as inconsistencies in gene symbols and heteromeric regulatory complexes representation. This enables large-scale comparative systems biology studies aimed to understand the common organizing principles and particular lifestyle adaptations of regulatory systems across bacteria and to implement those principles into future work such as the reverse engineering of GRNs.

## Availability and requirements

Abasy Atlas is available for web access at https://abasy.ccg.unam.mx. If you use any material from Abasy Atlas please cite properly. Use of Abasy Atlas and each downloaded material is licensed under a Creative Commons Attribution 4.0 International License. Permissions beyond the scope of this license may be available at jfreyre@ccg.unam.mx. **Disclaimer:** Please note that original data contained in Abasy Atlas may be subject to rights claimed by third parties. It is the responsibility of users of Abasy Atlas to ensure that their exploitation of the data does not infringe any of the rights of such third parties.

## List of abbreviations

GRN – Gene regulatory network

TFs – Transcription factors

TGs – Target genes

NDA – Natural decomposition approach

## Conflict of interest

The authors declare no conflicts of interest.

## Funding

This work was supported by the Programa de Apoyo a Proyectos de Investigación e Innovación Tecnológica (PAPIIT-UNAM), México [IN205918 to JAFG].

## Acknowledgments

We thank all anonymous reviewers for their positive and encouraging suggestions that helped to improve the manuscript. JMER was supported by an undergraduate fellowship from DGAPA-UNAM. He is also a PhD student from Programa de Doctorado en Ciencias Biomédicas, Universidad Nacional Autónoma de México (UNAM) and received fellowship 959406 from CONACYT.

58. Chauhan, R, Ravi, J, Datta, P, Chen, T, Schnappinger, D, et al. (2016) Reconstruction and topological characterization of the sigma factor regulatory network of Mycobacterium tuberculosis. Nat Commun 7: 11062.
59. Salgado, H, Peralta-Gil, M, Gama-Castro, S, Santos-Zavaleta, A, Muniz-Rascado, L, et al. (2013) RegulonDB v8.0: omics data sets, evolutionary conservation, regulatory phrases, cross-validated gold standards and more. Nucleic Acids Res 41: D203-213.

**Figures**

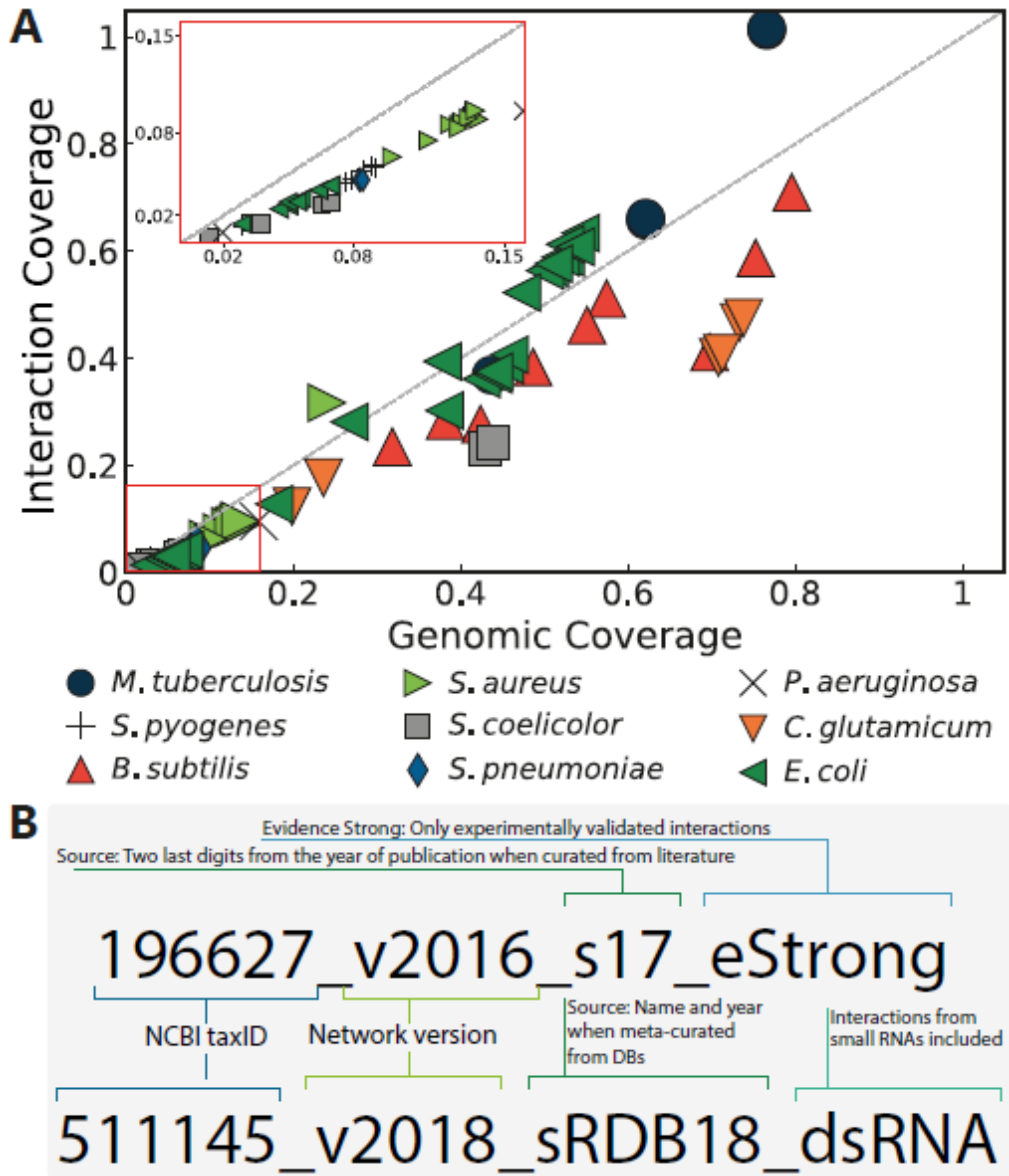

Figure 1. Abasy Atlas content. (**A**) Completeness measured as genomic and interaction coverage for the GRNs in Abasy, 76 networks covering 42 bacteria distributed in 9 species. (**B**) Examples

describing the format of the Abasy identifiers. The most complete *C. glutamicum* GRN (upper) filtered to contain only "strong" interactions, and the most recent, meta-curated *E. coli* GRN (lower).

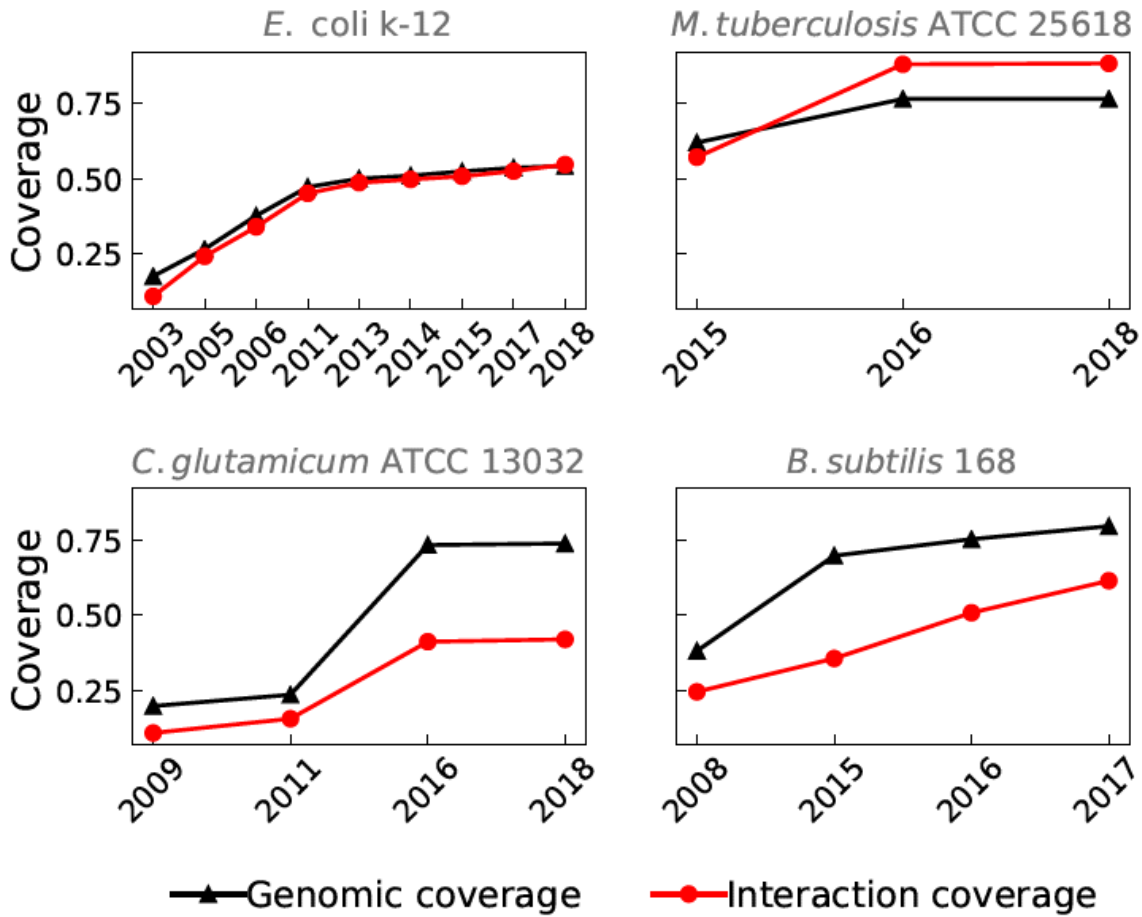

Figure 2. Historical snapshots for GRNs of model organisms. The completeness of the network can be measured as genomic coverage (fraction of the genome included in the GRN, black triangles) and interaction coverage (fraction of the known interactions relative to the complete network, red circles). It is evident that for some networks genomic coverage overestimates completeness as some networks may be classified as almost completed in terms of genomic coverage whereas many interactions are still missing. For instance, the GRN for *C. glutamicum* in 2016 is a meta-curation of the network from 2011 and a set of interactions curated in [13] including the *sigA* housekeeping sigmulon. On the other hand, the GRN for *M. tuberculosis* in 2016 is the most complete in terms of interaction coverage (97.7%) since it integrates the network from 2015 with novel interactions curated from the literature.

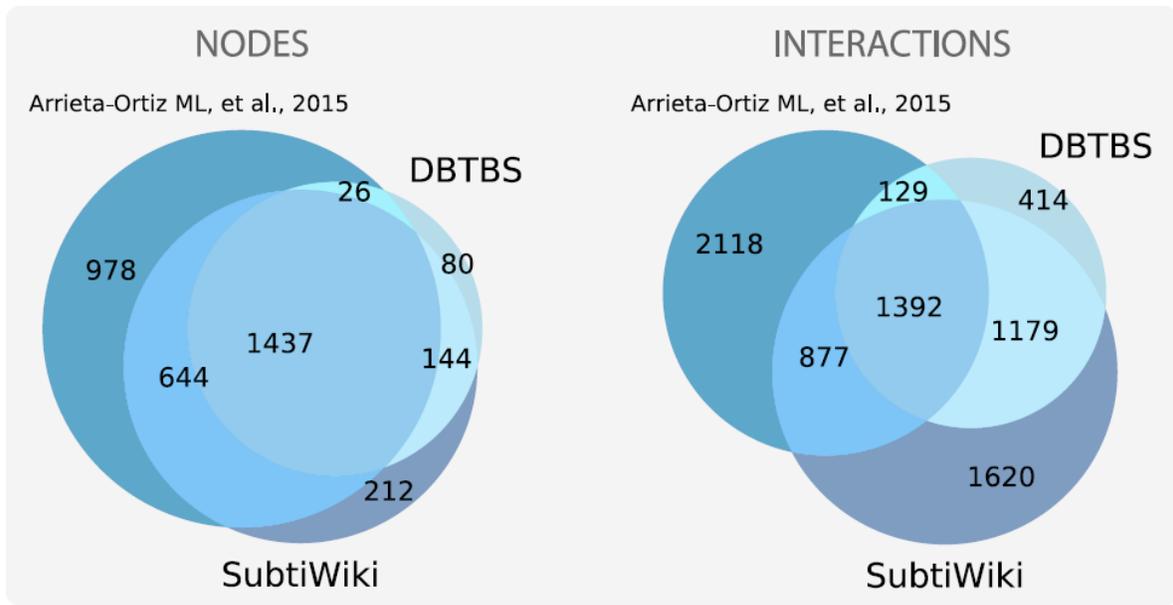

Figure 3. Complementary sources to reconstruct the meta-curated GRN for *B. subtilis*. A poor overlap is observed between the different sources used to reconstruct the meta-curated GRN for *B. subtilis*, mainly for interactions. This highlights the need for the meta-curation since the organism-specific databases do not fully cover each other nor the dataset not previously hosted in any database. Abasy provides homogeneous meta-curations integrating all the available information.

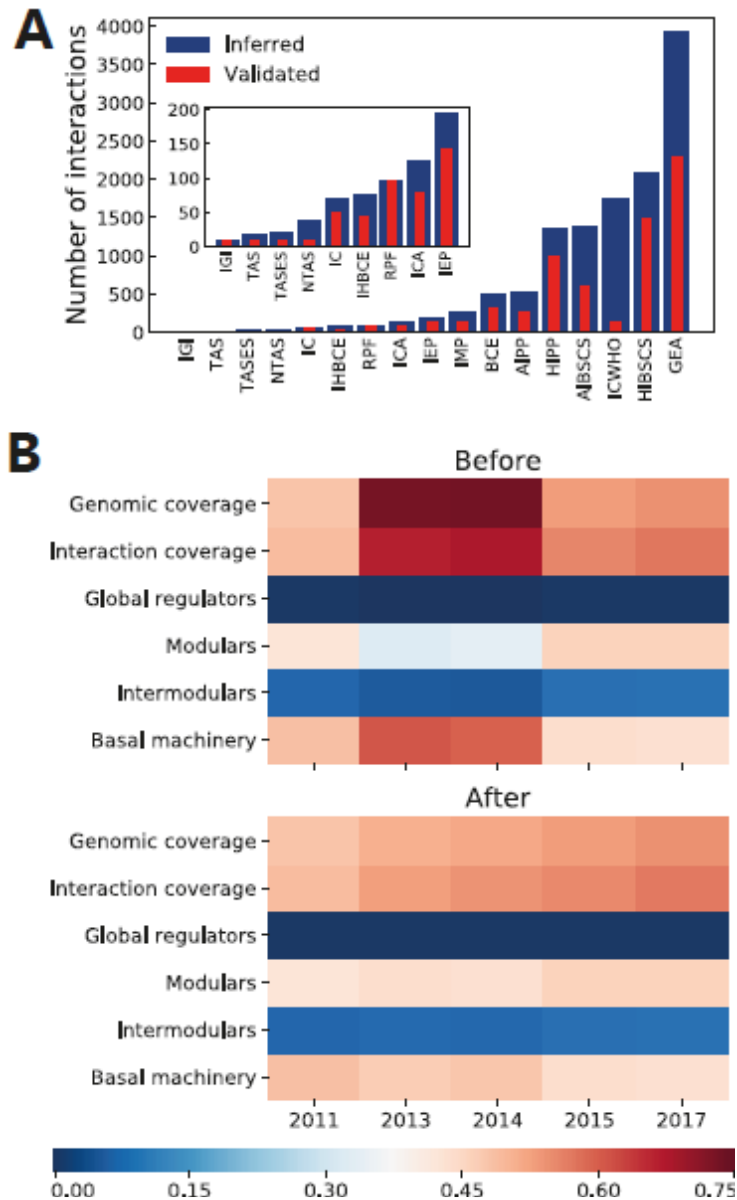

Figure 4. (A) Number of interactions identified by methods described as "weak" in [3] and how many of these interactions have been validated by "strong" evidence. IGI (inferred from genetic interaction), TAS (traceable author statement), TASES (traceable author statement to experimental support), NTAS (non-traceable author statement), IC (inferred by curator), IHBCE (inferred by a human based on computational evidence), RFP (RNA-polymerase footprinting), ICA (inferred by computational analysis), IEP (inferred from expression pattern), IMP (inferred from mutant phenotype), BCE (binding of cellular extracts), AIPP (automated inference of promoter position), HIPP (human inference of promoter position), AIBSCS (automated inference based on similarity to consensus sequences), ICWHO (inferred computationally without human oversight), HIBSCS (human inference based on similarity to consensus sequences), GEA (gene expression analysis) [59]. (B) Effect of removing spurious interactions through the meta-curation process. System-level elements (global regulators, modular, intermodular, and basal-machinery genes) values represent its fraction

from the total genes in the *E. coli* GRN historical snapshots before and after removal of interactions supported only by the ICWHO evidence.

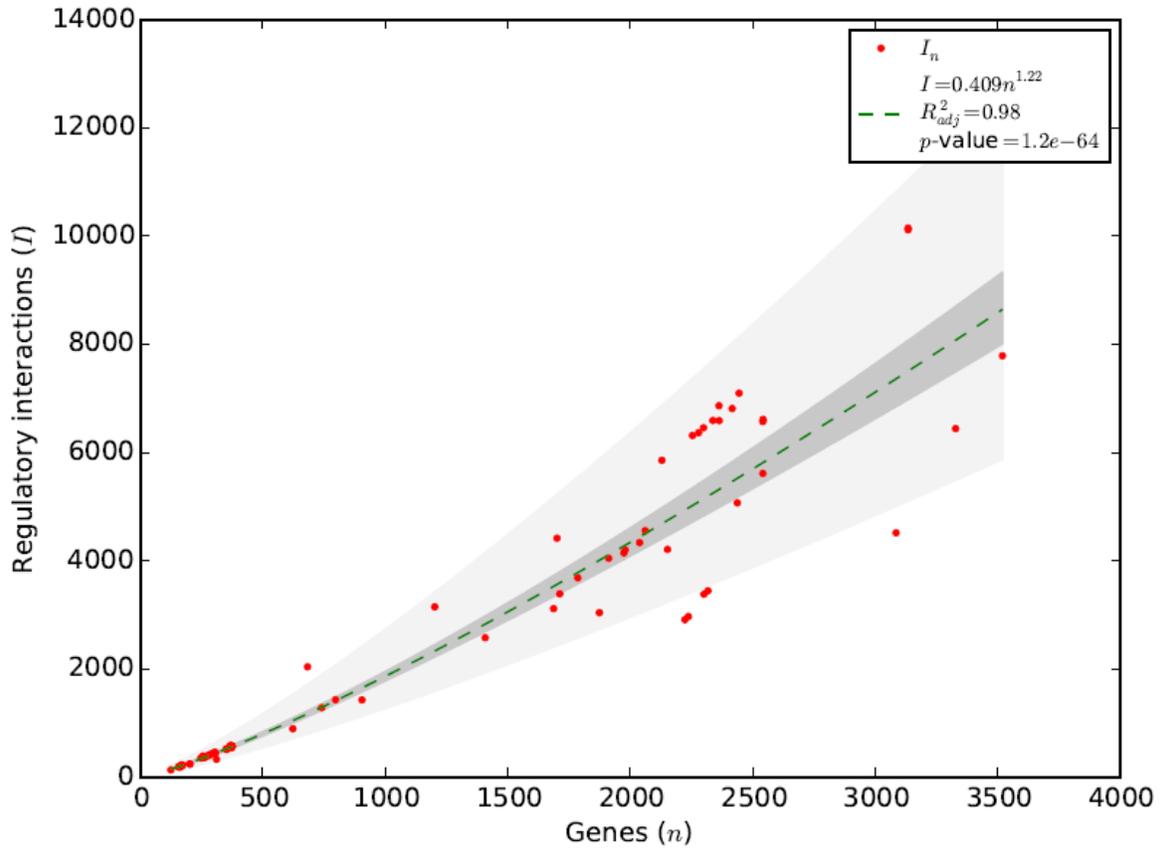

Figure 5. The constrained complexity of regulatory networks allows computing their total number of interactions. The number of interactions in the Abasy GRNs follows a power law with the number of genes as $I \sim n^{2-\gamma}$ ($R^2$ = 0.98), where $\gamma$ is the exponent of the power law found for the density of these networks. This power law may be used to compute the total number of interactions ($I_{total}$) in the regulatory network of an organism as $I_{total} \sim$ (genome size)$^{2-\gamma}$.

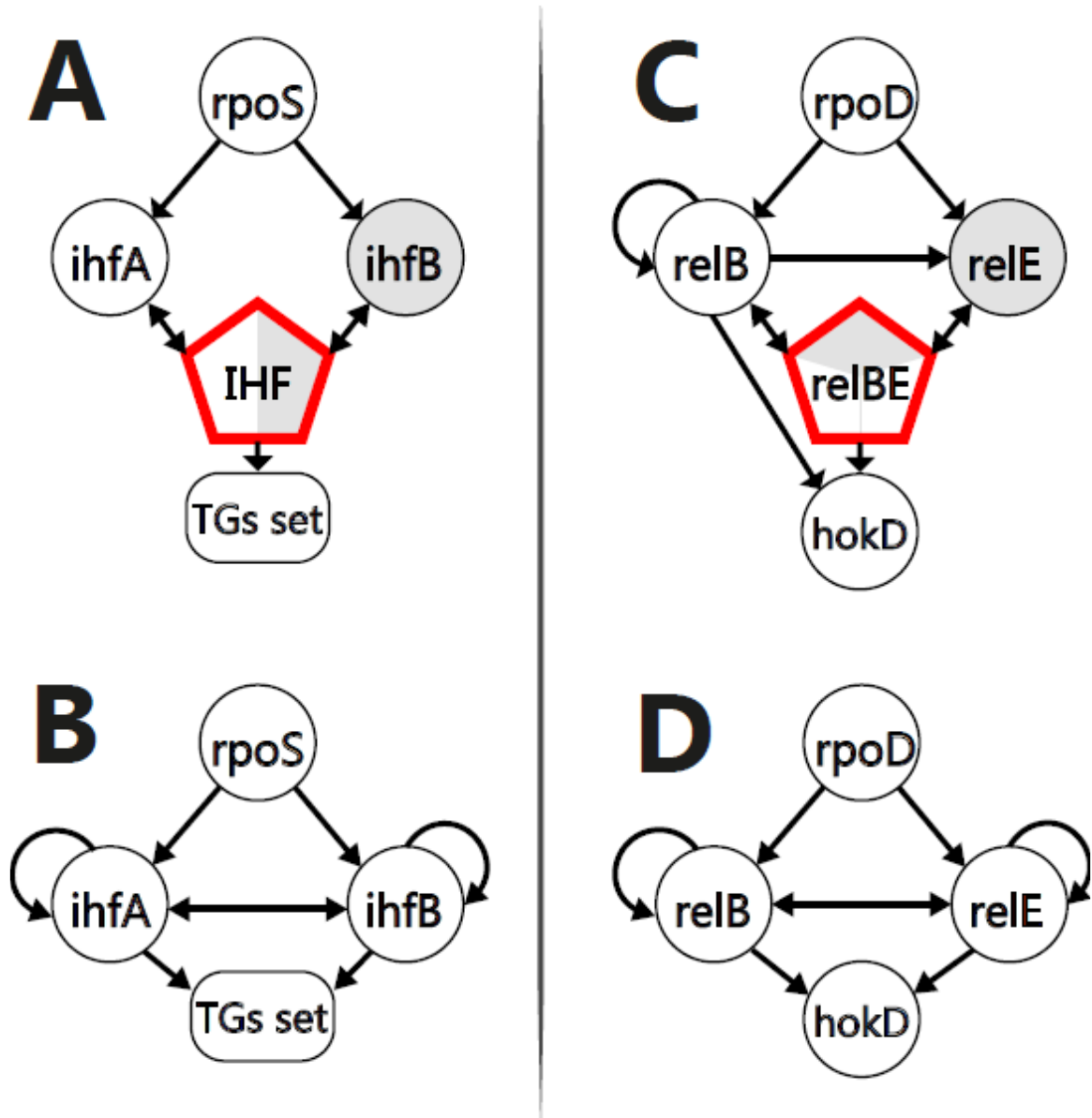

Figure 6. Homogeneous network representation. Heteromeric-complex-base gene representation for IHF (**A**) and RelBE (**C**). Misrepresentation of gene expression regulation where heteromeric protein complexes are involved for IHF (**A**) and RelBE (**D**) systems. RelB can regulate itself as a homomeric-complex, and as a heteromeric-complex with *relE* (**C**). Besides, *relE* can regulate neither its transcription nor RelB transcription on its own, as could be misinterpreted from (**E**). This same misrepresentation is observed for the IHF complex where neither of the subunits has regulatory activity as a homomeric complex.

**Supplementary data**

Supplementary information is available online.

2626

# Supplementary information for

**Abasy Atlas v2.2: The most comprehensive inventory of meta-curated, historical and up-to-date, bacterial regulatory networks, their completeness and system-level characterization**

Juan M. Escorcia-Rodríguez[1], Andreas Tauch[2], and Julio A. Freyre-González[1,*]

[1]Regulatory Systems Biology Research Group, Laboratory of Systems and Synthetic Biology, Center for Genomics Sciences, Universidad Nacional Autónoma de México. Av. Universidad s/n, Col. Chamilpa, 62210. Cuernavaca, Morelos, México

[2]Centrum für Biotechnologie (CeBiTec). Universität Bielefeld, Universitätsstraße 27, 33615. Bielefeld, Germany

**\*Corresponding author:** jfreyre@ccg.unam.mx (JAFG)

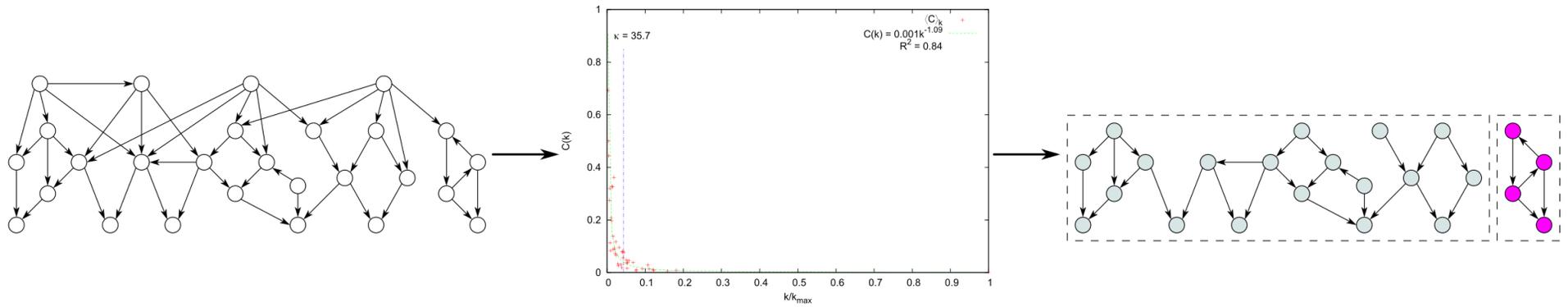

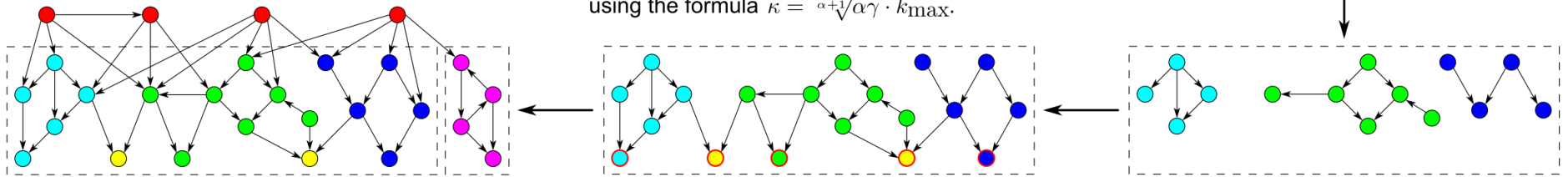

A regulatory network is represented as a directed graph. Each gene is a node, while arcs between them stand for regulatory interactions.

Global regulators are identified by the κ-value, which defines an equilibrium point between two contradictory behaviors: hubness (high out-degree, low clustering) and modularity (low out-degree, high clustering). First, the clustering coefficient distribution, $C(k)$, is computed. Then, given $C(k) = \gamma k^{-\alpha}$, the κ-value is calculated by using the formula $\kappa = {}^{\alpha+1}\!\sqrt{\alpha\gamma} \cdot k_{\max}$.

Identification of functional modules. Global regulators (connectivity > κ-value) and their links are removed from the network, thus naturally revealing the modules (isolated islands) and the basal machinery genes (single isolated nodes not encoding for regulators, not shown here for the sake of clarity).

Finally, global regulators (red nodes) and their links are added back, thus reconstructing the original network but additionally revealing its hidden diamond-shaped three-layered architecture and its systems-level components: global regulators, modules, basal machinery and intermodular genes.

Identification of intermodular genes. All the removed non-regulators-encoding genes (red-outlined nodes) and their interactions are added back following a rule: if gene G is regulated by genes belonging to two or more submodules, then G is classified as intermodular (yellow nodes), else G is added to the same submodule than its regulators.

Identification of pre-submodules composing the megamodule. The megamodule is isolated and all the non-regulator-encoding genes and their links are removed, disaggregating it into isolated islands. Each island is identified as a pre-submodule.

**Supplementary Figure 1.** Methodology of the natural decomposition approach (NDA).

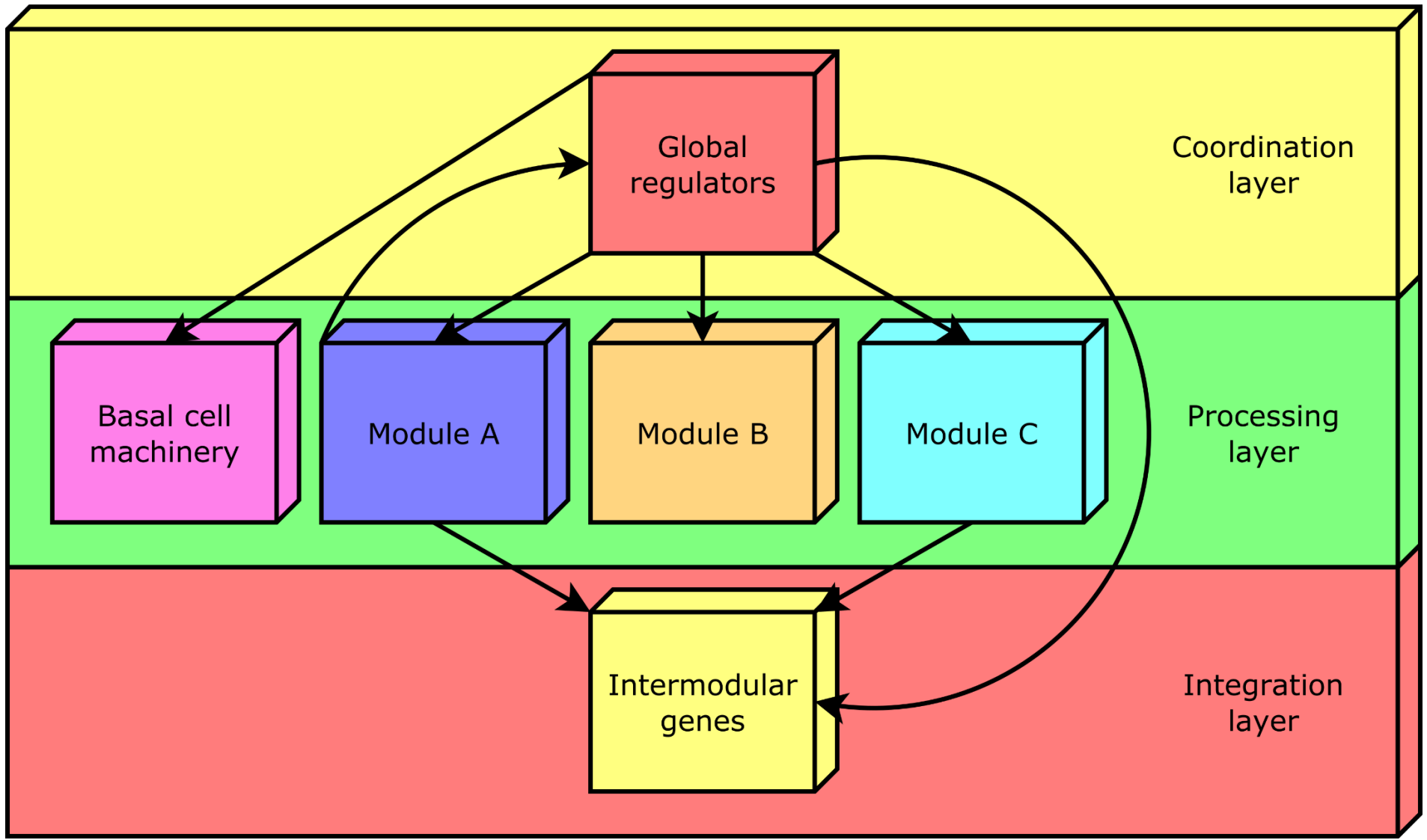

**Supplementary Figure 2.** The functional architecture unveiled by the NDA is a diamond-shaped, three-tier hierarchy, exhibiting some feedback between processing and coordination layers, which is shaped by four classes of system-level elements: global regulators, locally autonomous modules, basal machinery, and intermodular genes.

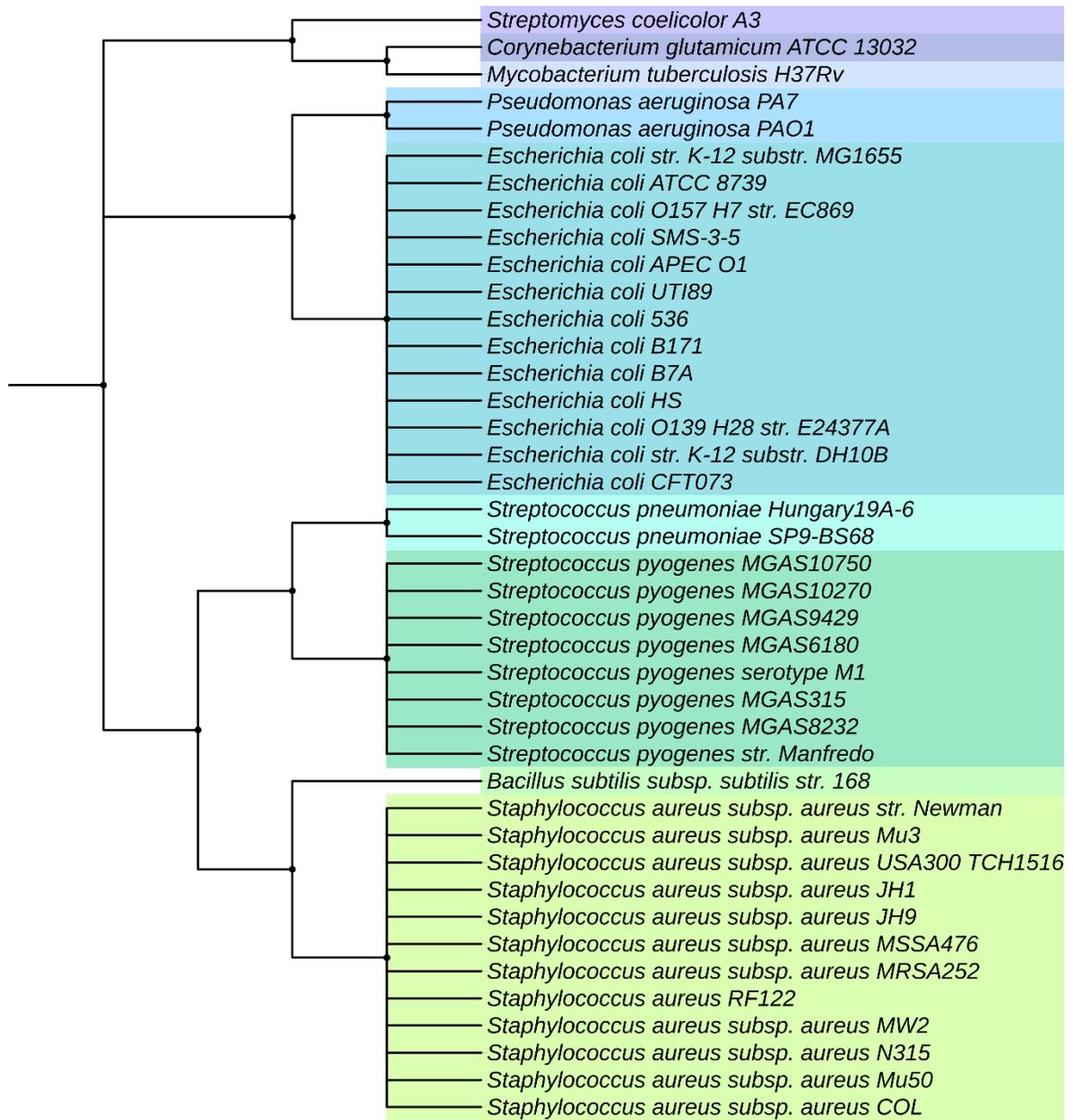

**Supplementary Figure 3.** Abasy integrates gene regulatory interactions for 9 species and 41 strains including model organisms.

| DB abbreviation | PMID | Reference |
|---|---|---|
| Bsub15 | 26577401 | [1] |
| CRN12 | 22080556 | [2] |
| Cglu17 | 27829123 | [3] |
| DBTBS08 | 17962296 | [4] |
| Mtuber11 | 21818301 | [5] |
| Mtuber12 | 22737072 | [6] |
| Mtuber15 | 25581030 | [7] |
| Mtuber16 | 27029515 | [8] |
| Paeru11 | 22587778 | [9] |
| RDB01 | 11125053 | [10] |
| RDB04 | 14681419 | [11] |
| RDB06 | 16381895 | [12] |
| RDB11 | 21051347 | [13] |
| RDB13 | 23203884 | [14] |
| RDB16 | 26527724 | [15] |
| RDB18 | 30395280 | [16] |
| RTB13 | 23547897 | [17] |
| SW16 | 26433225 | [18] |
| SW18 | 29788229 | [19] |

**Supplementary Table 1.** Name abbreviation and year used as code for the source field in the network ID, and the PMID of the article.

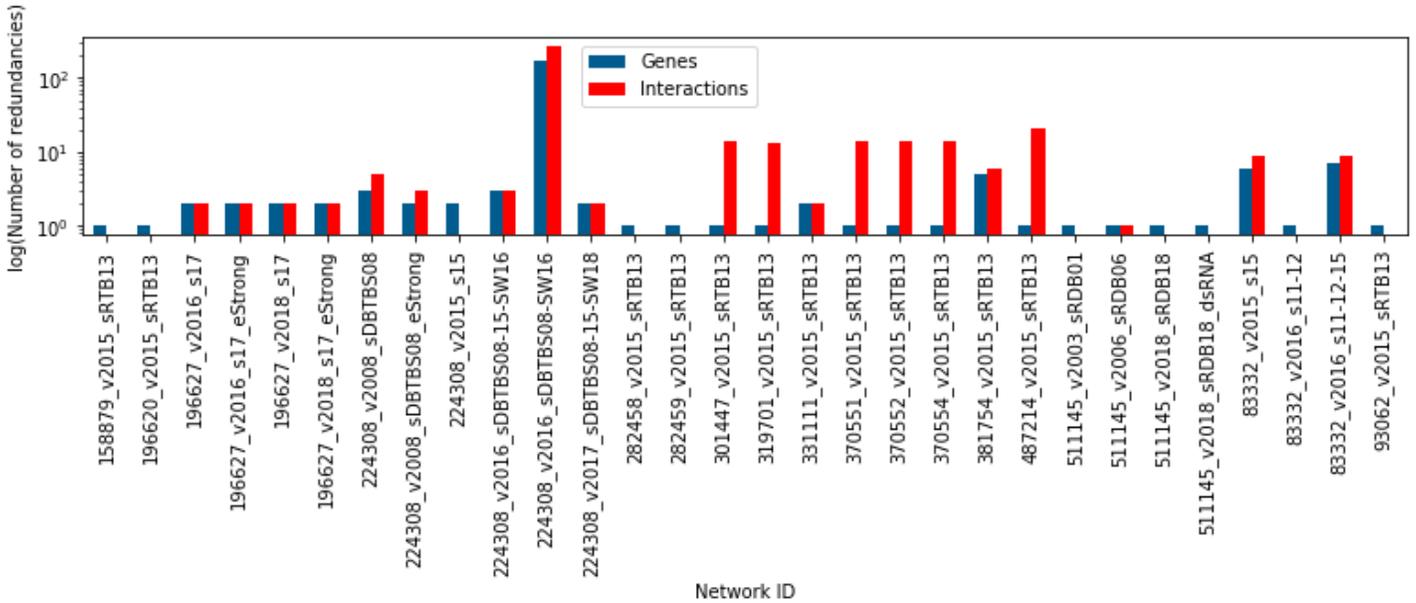

**Supplementary Figure 4.** Number of redundant genes and interactions removed due to the synonyms resolution in the meta-curated GRNs. A total of 223 nodes and 412 interactions are removed.

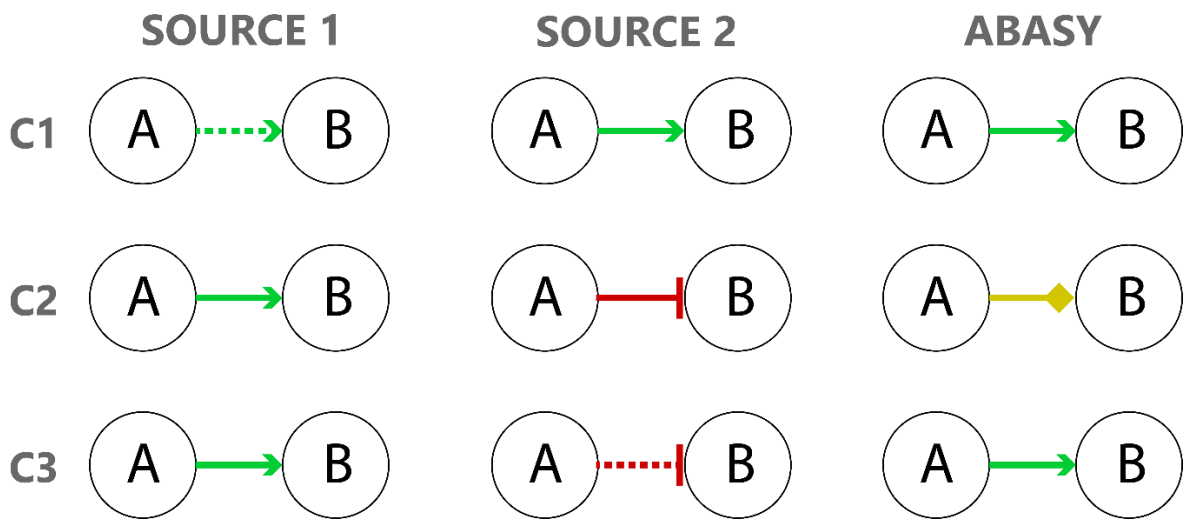

**Supplementary Figure 5.** Information preserved when consolidating regulatory interactions from different sources. **C1:** In case of same effect but different evidence level, the interaction shared and the "strong" evidence is conserved. **C2:** In case of different effects and the same evidence level, both effects are conserved in a single dual interaction to avoid redundancy. **C3:** In case of both attributes are different, only the "strong" interaction is conserved.

**Supplementary Figure 6.** Filtered GRNs to contain only "strong" interactions van be also downloaded through the "Downloads" page where you can select the GRN you want as well as the additional data such as gene information ad modules annotation.

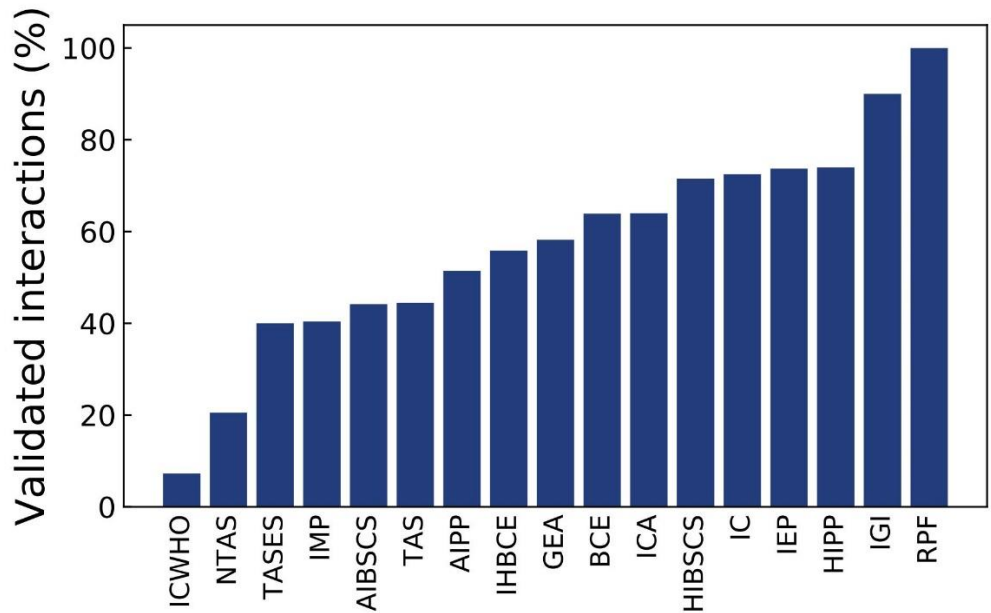

**Supplementary Figure 7.** Interactions inferred by a methodology classified as "weak" that has been validated with a "strong" experiment.

![Screenshot of RegulonDB interface showing Escherichia coli str. K-12 substr. MG1655 [2018, RDB18, Strong] All genes/regulatory complexes table with columns Name, Biological entity, Product description, NDA class. Rows include RelB-RelE (Regulatory complex, –, 1), YefM-YoeB (Regulatory complex, –, 68), accA (Gene, acetyl-CoA carboxyltransferase, α subunit, 60.7), accB (Gene, biotin carboxyl carrier protein, 60.7), accC (Gene, AccC, 60.7), accD (Gene, acetyl-CoA carboxyltransferase, β subunit, 60.7), aceA (Gene, isocitrate lyase monomer, Intermodular), aceB (Gene, malate synthase A, Intermodular), aceE (Gene, subunit of E1p component of pyruvate dehydrogenase complex, 60.8), aceF (Gene, AceF, 60.8), aceK (Gene, AceK, Intermodular), acnA (Gene, aconitate hydratase 1, 60.2), acnB (Gene, bifunctional aconitate hydratase 2 and 2-methylisocitrate dehydratase, Basal machinery), acpP (Gene, apo-[acyl carrier protein], Basal machinery), acpS (Gene, AcpS, Basal machinery).]

**Supplementary Figure 8.** Regulatory complexes can be identify using the "Biological entity" column and the Gene information file available at the "Downloads" page (see **Supplementary Figure 2)**. This makes possible to convert the GRNs to contain only gene-gene interactions when needed.

| Regulatory network | Version | Network genomic coverage | Constrained complexity model / Network completeness | Data sources (PMID) | NDA-predicted system-level components | | | | |
|---|---|---|---|---|---|---|---|---|---|
| | | | | | Global regulators | Modules | Modulars | Basal machinery | Intermodulars |
| ▶ *Mycobacterium tuberculosis* (5 items) | | | | | | | | | |
| ▶ *Bacillus subtilis* (9 items) | | | | | | | | | |
| ▶ *Escherichia coli* (30 items) | | | | | | | | | |
| ▼ *Corynebacterium glutamicum* (6 items) | | | | | | | | | |
| ▼ strain ATCC 13032 / DSM 20300 / JCM 1318 / LMG 3730 / NCIMB 10025 (NCBI taxid: 196627) (6 items) | | | | | | | | | |
| ▶ Indirectly and directly experimentally validated interactions (Weak + Strong evidences) (2 items) | | | | | | | | | |
| ▼ Directly experimentally validated interactions (Strong evidences) (4 items) | | | | | | | | | |
| 196627_v2018_s17_eStrong (NCBI TaxID: 196627) Global properties | 2018 | 71.3% (2237 / 3138) | 39.5% (2969 / 7516) [-2.76%, 2.97%] | 27829123 | 4 0.18% | 58 | 510 22.8% | 1675 74.9% | 48 2.15% |
| 196627_v2016_s17_eStrong (NCBI TaxID: 196627) Global properties | 2016 | 70.8% (2223 / 3138) | 38.7% (2911 / 7516) [-2.71%, 2.91%] | 27829123 | 4 0.18% | 56 | 459 20.6% | 1713 77.1% | 47 2.11% |
| 196627_v2011_s17_eStrong (NCBI TaxID: 196627) Global properties | 2011 | 23.6% (741 / 3138) | 17.1% (1283 / 7516) [-1.19%, 1.28%] | 27829123 | 21 2.83% | 60 | 304 41% | 414 55.9% | 2 0.27% |
| 196627_v2009_sCRN12_eStrong (NCBI TaxID: 196627) Global properties | 2009 | 19.9% (623 / 3138) | 11.9% (895 / 7516) [-0.83%, 0.89%] | 22080556 | 23 3.69% | 47 | 205 32.9% | 393 63.1% | 2 0.32% |
| ▶ *Staphylococcus aureus* (13 items) | | | | | | | | | |
| ▶ *Pseudomonas aeruginosa* (2 items) | | | | | | | | | |
| ▶ *Streptococcus pyogenes* (8 items) | | | | | | | | | |

**Supplementary Figure 9.** From the "Browse" page you can identify the species and strain of interest, as well as the confidence level you need. Also, you will find additional information such as the GC and IC, data sources and fraction of the systems-level components predicted by the NDA. Clicking the hyperlink in the data sources PMID will open the PubMed page of the article for the data source.

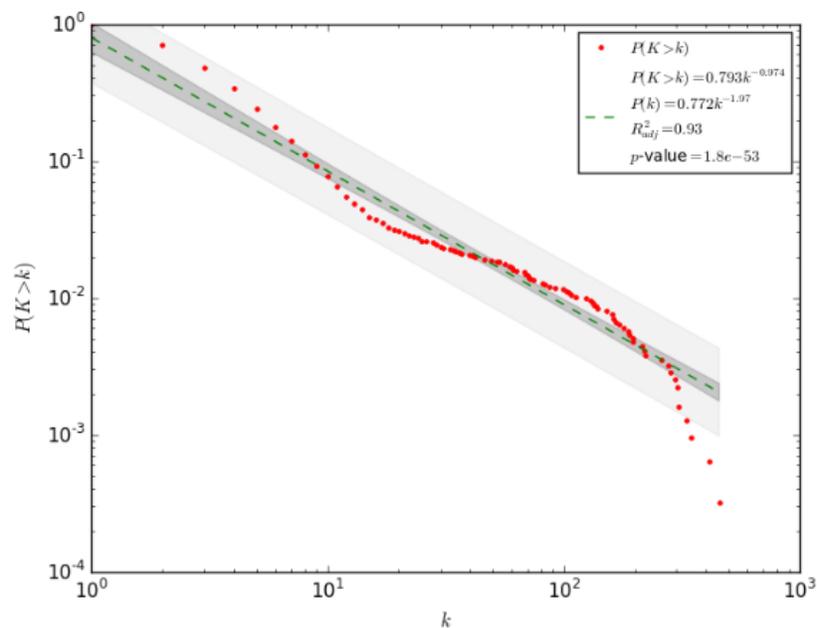

*Mycobacterium tuberculosis* (strain ATCC 25618 / H37Rv) [2018, 11-12-15-16, Weak + Strong]

**Network global properties**

| | | |
|---|---|---|
| Regulators ($k_{out} > 0$) | – | 179 (5.7%) |
| Structural genes ($k_{out} = 0$) | – | 2955 (94.3%) |
| Undirected regulatory links | – | 10119 |
| Directed regulatory interactions | – | 10147 |
| Self-regulations | – | 104 (58.1%) |
| Maximum out-connectivity | – | 452 (14.4%) |
| Network density | – | 0.00204 |
| Weakly connected components | – | 3 |
| Genes in the giant component | – | 3128 (99.8%) |
| Feedforward circuits | – | 3994 |
| Complex feedforward circuits | – | 700 |
| 3-Feedback loops | – | 63 |
| Average path length | – | 3.24 |
| Network diameter | – | 6 |
| Average clustering coefficient | – | 0.13008 |
| $P(k)$ | – | $0.772 \cdot k^{-1.97}$ |
| $R^2_{adj}$ | – | 0.93 |

Plot legend:
- $P(K>k)$
- $P(K>k) = 0.793 k^{-0.974}$
- $P(k) = 0.772 k^{-1.97}$
- $R^2_{adj} = 0.93$
- $p\text{-value} = 1.8e{-}53$

**Supplementary Figure 10.** Statistical and structural properties characterizing the GRN of interest.

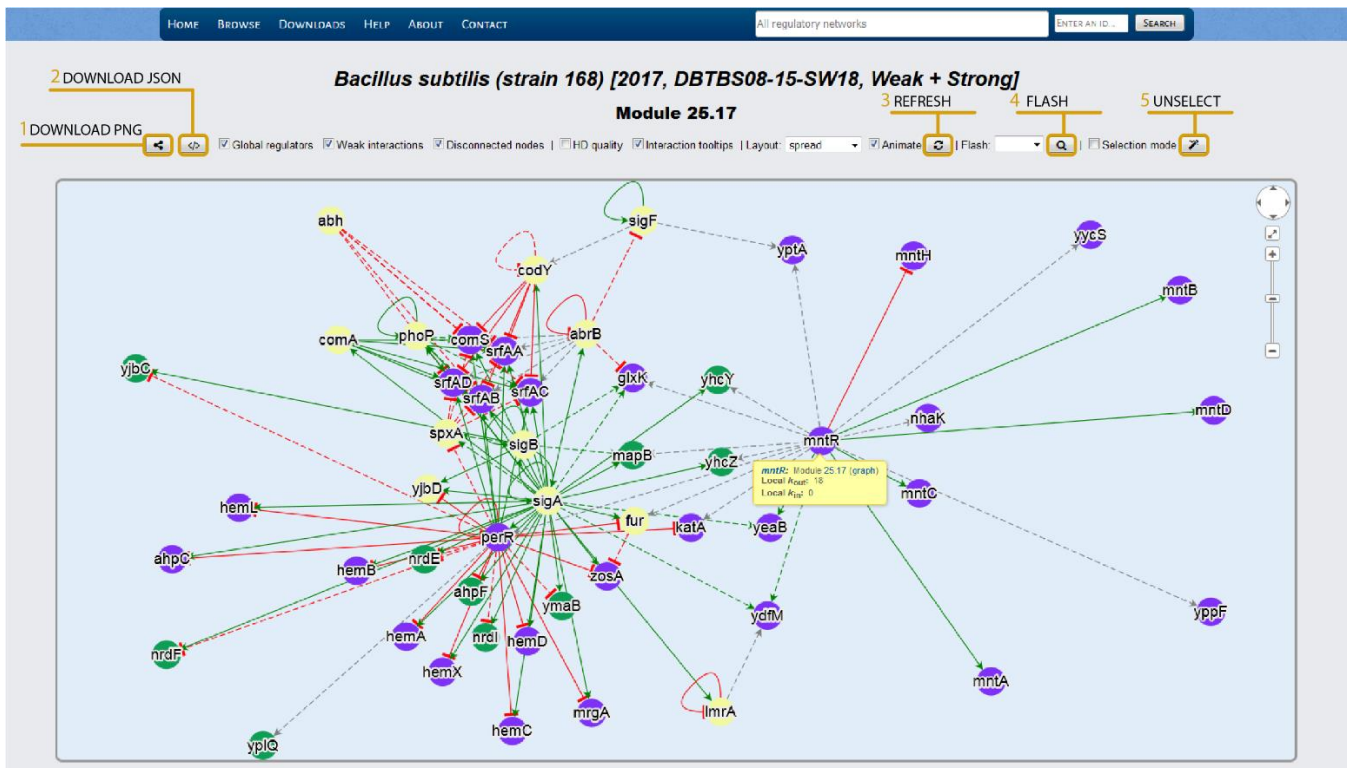

**Supplementary Figure 11.** On the interactive panel you can download the displayed network as an image in PNG format with transparent background (1). Also, you can download the JSON file (2) to import into Cytoscape and customize the style of the network. Check/uncheck the boxes to remove global regulators, "weak" interactions, disconnected genes, interaction tooltips and to display the network in high definition. One the desired boxes are checked, click the refresh button (3) to visualize the changes. Select one gene from the alphabetically sorted "Flash" list and click the flash button (4) to easily identify the gene of interest from the network. Click on the nodes to select them and drag the mouse to change the position of the selected nodes. To deselect all nodes, simply click the unselect button (5).